\documentclass[12pt]{iopart}
 
 \usepackage{iopams}  
\usepackage{graphicx}
\usepackage{multirow}
\usepackage{times}
\usepackage{amssymb}
\usepackage{dcolumn}%
\usepackage{bm}%
\usepackage[dvipdfmx]{hyperref}%
\usepackage{upgreek}
\usepackage{wasysym}
\usepackage{color}

\begin{document}

\title[Anisotropy of the monomer random walk]{Anisotropy of the monomer random walk in a polymer melt: local-order and connectivity effects}

\author{S.\@ Bernini${}^1$, D.\@ Leporini${}^2$}

\address{${}^1$ Dipartimento di Fisica ``Enrico Fermi'', 
Universit\`a di Pisa, Largo B.\@Pontecorvo 3, I-56127 Pisa, Italy}

\address{${}^2$ Dipartimento di Fisica ``Enrico Fermi'', 
Universit\`a di Pisa, Largo B.\@Pontecorvo 3, I-56127 Pisa, Italy and IPCF-CNR, UOS Pisa, Italy}
\ead{dino.leporini@unipi.it}

\begin{abstract}
The random walk of a bonded monomer  in a polymer melt is anisotropic due to 
local order and bond connectivity. We investigate both effects by 
molecular-dynamics simulations on melts of fully-flexible linear chains ranging 
from dimers ($M=2$) up to  entangled  polymers ($M=200$). The corresponding 
atomic liquid is also considered as reference system. To disentangle the 
influence of the local geometry and the bond arrangements, and reveal their 
interplay, we define suitable measures of the anisotropy emphasising either the 
former or the latter aspect. Connectivity anisotropy, as measured by the 
correlation between the 
initial bond orientation and the direction of the subsequent monomer 
displacement, shows a slight {\it enhancement} due to the local order at  times 
shorter than the structural relaxation time.
At intermediate times - when the monomer displacement is comparable to the bond 
length - a pronounced peak   and then decays slowly as $t^{-1/2}$, becoming 
negligible when  the displacements is as large as about five bond lengths, i.e. 
about four monomer diameters or three Kuhn lengths. Local-geometry anisotropy, 
as measured by the correlation between the initial orientation of 
a characteristic axis of the Voronoi cell and the subsequent monomer dynamics, 
 is affected at shorter times than the structural relaxation time by the cage 
shape with {\it antagonistic disturbance} by the connectivity. Differently, at 
longer times, the connectivity {\it favours} the persistence of the 
local-geometry anisotropy which vanishes when the monomer displacement exceeds 
the bond length. Our results strongly suggest that the sole consideration of 
the local order is not enough to understand the microscopic origin of the 
rattling amplitude of the trapped monomer in the cage of the neighbours.
\end{abstract}

%
%
%
%

\section{Introduction}

{The relation between the structure and the dynamics is a key problem in liquid-state 
\cite{HansenMcDonaldIIIEd,GotzeBook,BerthierBiroliRMP11} and polymer \cite{PaulSmithRepProgrPhys04,sim} physics. 
In particular, in a liquid of particles the random walk of the constituents is 
affected at short times (or length scales comparable to the particle diameter) 
by the local order and the displacement is anisotropic, differing by the one of 
a brownian particle in an homogeneous liquid  
\cite{TorquatoStilliPRE02,AsteJPCM05,AstePRE05}.
In a polymer melt, the anisotropy of the bonded-monomer 
displacement is  also contributed by the bond connectivity 
\cite{DoiEdwards,Rubinstein,Strobl}. While the local-order anisotropy (LOA) is 
expected to be stronger at shorter times, the connectivity anisotropy (COA) is 
anticipated to be important at both short and long times, since it involves the 
connected particles on different length scales 
\cite{DoiEdwards,Rubinstein,Strobl}. The interplay between the connectivity and 
the local order is, at least in part, antagonistic 
\cite{LocalOrderJCP13}. In fact, the connectivity introduces a 
length scale, the bond length, which perturbs the local order with 
characteristic length scale given by the particle size, in general not 
commensurate with the bond length.

Our interest in the relation between particle displacement and local ordering is motivated by a number of arguments. 
The main motivation is the pursuit  of the microscopic origin of the universal correlation between the mean square 
amplitude of the cage rattling (related to the Debye-Waller factor) and the relaxation and transport, as found in 
simulations of polymers \cite{OurNatPhys,lepoJCP09,Puosi11}, binary atomic mixtures \cite{lepoJCP09,SpecialIssueJCP13}, 
colloidal gels \cite{UnivSoftMatter11} and antiplasticized polymers 
\cite{DouglasCiceroneSoftMatter12,DouglasStarrPNAS2015},
and supported by the experimental data concerning several glassformers in  a wide fragility range ($20 \le m \le 191$) 
\cite{OurNatPhys,UnivPhilMag11,OttochianLepoJNCS11,SpecialIssueJCP13,CommentSoftMatter13}. 
From this respect, the local structure due to the first neighbours was recently found to correlate poorly with the rattling amplitude in the cage of the closest neighbours, and the structural relaxation, in liquids of linear trimers  \cite{VoroBinarieJCP15,VoronoiBarcellonaJNCS14} and atomic mixtures  \cite{VoroBinarieJCP15}. On the other hand, extended modes ranging up to about the fourth shell do 
correlate with the rattling amplitude in the cage and the structural relaxation \cite{PuosiLepoJCPCor12,PuosiLepoJCPCor12_Erratum}. These two complementary findings are fully consistent with Berthier and Jack who concluded that the influence of structure on dynamics is weak on short length scale and becomes much stronger on long length scale \cite{BerthierJackPRE07}.
On a more general grounds, several approaches  suggest that structural aspects matter in the dynamics of glassforming systems. This includes the 
Adam-Gibbs derivation of the structural relaxation \cite{AdamGibbs65,DudowiczEtAl08} - built on the thermodynamic notion 
of the configurational entropy \cite{GibbsDiMarzio58} -, the mode- coupling theory \cite{GotzeBook} and extensions 
\cite{SchweizerAnnRev10}, the random first-order transition theory \cite{WolynesRFOT07}, the frustration-based approach 
\cite{TarjusJPCM05}, as well as the so-called  elastic models  
\cite{Dyre06,Nemilov06,StarrEtAl02,LemaitrePRL14,GranatoFragilityElasticityJNCS02,WyartPNAS2013,NovikovEtAl05,Novikov04,
SchweizerElastic1JCP14,SchweizerElastic2JCP14,DouglasStarrPNAS2015,Puosi12,DyreWangGpJCP12,ElasticoEPJE15,
BerniniElasticJPolB15}
in that the modulus is set by the  arrangement of the particles in mechanical equilibrium and their mutual interactions 
\cite{Dyre06,Puosi12}. It was concluded that the proper inclusion of many-body static correlations in theories of the 
glass transition appears crucial for the description of the dynamics of fragile glass formers \cite{coslovichPRE11}.
The search of a link between structural ordering and slow dynamics motivated several studies in liquids 
\cite{NapolitanoNatCom12,EdigerDePabloNatMat13,BarbieriGoriniPRE04,ReichmannCoslovichLocalOrderPRL14,RoyallNatCom15}
colloids \cite{StarrWeitz05,TanakaNatMater08,TanakaNatCom12} and polymeric systems 
\cite{StarrWeitz05,DePabloJCP05,GlotzerPRE07,LasoJCP09,BaschnagelEPJE11,MakotoMM11,LariniCrystJPCM05}.

Here, we present a detailed  study of both LOA and COA anisotropies of the 
bonded monomers by molecular-dynamics (MD) simulation of a  polymeric melt  of 
linear chains and an atomic liquid (as reference system in one particular 
state). As to the polymer melt, particular attention is devoted to temperature 
and the chain length which is changed from oligomers (trimers, $M=3$) up to 
entangled systems ($M=200$).

LOA is characterized by the correlation between the 
initial  shape of the cage of the closest neighbours and the direction of the subsequent monomer displacement. We are inspired by a seminal work by Rahman in an atomic liquid \cite{Rahman66}, studying the {\it directional} 
correlations between the particle dynamics of the trapped particle in the cage and the  position of the centroid 
{\bf C} of the  vertices of the associated Voronoi polyhedron (VP). The interest relies on the fact that the VP vertices 
are located close to the voids between the particles and thus mark the weak spots of the cage. It has been shown in 
simulations of atomic liquids \cite{Rahman66} and experiments on granular matter \cite{DouglasAspehricity} that the  
particle initially moves towards the centroid, so that cage rattling and VP geometry are correlated at {very short} 
times. We are not aware of extensions to {\it molecular} liquids where COA is present.

COA is quantified by  the correlation between the initial bond orientation and the direction of the monomer 
displacement. 
}

The paper is organized as follows. In 
Sec.\;\ref{numerical} the molecular model and the MD algorithms are presented. The results are discussed 
in Sec.\;\ref{resultsdiscussion} and the conclusions are summarized in Sec.\ref{concl}.

\section{Methods}
\label{numerical}

A coarse-grained model of a melt of $N_c$ linear fully-flexible polymer chains with $M$ monomers per  chain is 
considered. Full flexibility is ensured by the absence of both torsional or 
bending potentials hindering the bond orientations. We set $M=2,3,5,10,15,30, 
100, 200$. Entanglements are expected for lengths exceeding $M_e$ with $M_e$ 
estimated, according to different methods, as $\simeq 70$ \cite{EveraersKremerGrestJPolSci05} $\simeq 74$ 
\cite{KremerGrestEntangleEPL00} and $\simeq 80$ \cite{HoyPRE09}. Non-bonded monomers at a distance $r$ interact via a 
truncated Lennard-Jones (LJ) potential $U_{LJ}(r) =\varepsilon \left [ \left (\sigma^*/ r \right)^{12 } - 2\left 
(\sigma^*/r\right)^6 \right]+U_{cut}$ { for $r< r_c=2.5\,\sigma$} and zero otherwise, where $\sigma^*=\sqrt[6]{2} \, 
\sigma$ is the position of the potential minimum with depth $\varepsilon$. The value of the constant $U_{cut}$ is chosen 
to ensure that $U_{LJ}(r)$ is continuous at $r = r_c$. The bonded monomers interact by a potential which is the sum of 
the LJ potential and the FENE (finitely extended nonlinear elastic) potential $ U^{FENE}(r)=-{1/2} \;kR_0^2 \; 
\ln\left(1-r^2/R_0^2\right)$ where $k$ measures the magnitude of the interaction and $R_0$ is the maximum elongation 
distance \cite{sim,VoronoiBarcellonaJNCS14}. The parameters $k$ and $R_0$ have been set to $30 \, \varepsilon  / 
\sigma^2 $ and $ 1.5\,\sigma $ respectively \cite{GrestPRA33}. The resulting bond length is $b=0.97\sigma$ within a few 
percent. All quantities are in reduced units: length in units of $\sigma$, temperature in units of $\varepsilon/k_B$ 
(with $k_B$ the Boltzmann constant) and time $\tau_{MD}$ in units of $\sigma \sqrt{m / \varepsilon}$ where $m$ is the 
monomer mass. We set $m = k_B = 1$. We investigate states with number density $\rho=1.086$, temperature $T=1$ and 
$(N_c,M)$ pairs: (1000, 2), (667, 3), (400, 5), (200, 10), (134, 15), (67, 30), (20, 100) and (60, 200). We also 
investigate states with  $T=0.9,0.8,0.7,0.63,0.6$ for the pair (667, 3). { In order to investigate the role of the 
connectivity we simulate an atomic Lennard-Jones liquid of $8000$ atoms at density $\rho=1.086$ and temperature $T=1.5$ and a 
molecular 
system at the same density and temperature with pair (2667, 3)}. Periodic boundary conditions are used. $NVT$ 
ensemble (constant number of particles, volume and temperature) has been used for equilibration runs, while $NVE$ 
ensemble (constant number of particles, volume and energy) has been used for 
production runs of a given state point with time step $3 \cdot 10^{-3}$ 
\cite{allentildesley}. The samples were equilibrated in lapses of 
time as long as, at least, three times the longest relaxation time, i.e. the 
average reorientation time of the chain $\tau_r$.  The simulations are carried 
out using LAMMPS molecular dynamics software 
(http://lammps.sandia.gov) \cite{PlimptonLAMMPS}.

\begin{figure}[t]
\begin{center}
\includegraphics[width=0.5\linewidth]{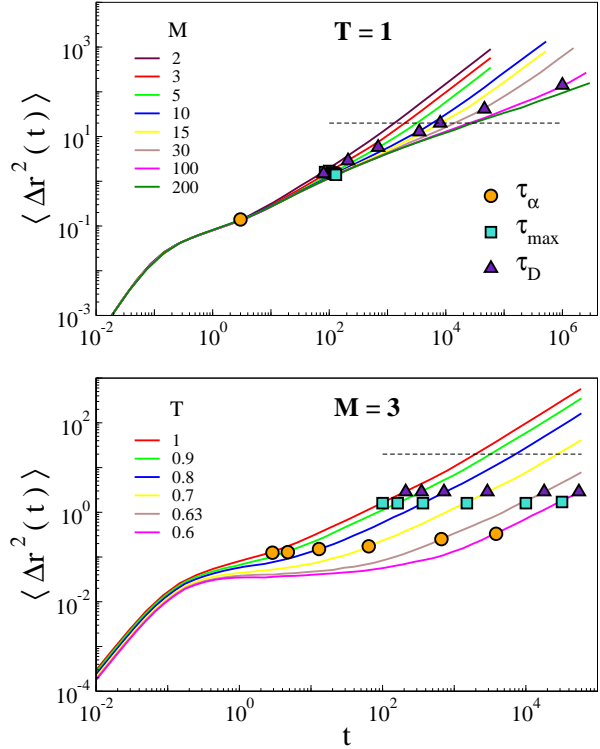}
\end{center}
\caption{Monomer MSD  for different chain lengths at $T=1$ (top) and for trimers at different temperatures (bottom). The 
orange circles mark the structural relaxation time $\tau_\alpha$. The turquoise squares mark the time  $\tau_{max}$ 
where the COA function $\widehat C(t)$, Eq.\ref{tetabond}, is maximum (see Fig.\ref{tetab}). The violet 
triangles mark the onset time $\tau_D$ of the diffusion regime defined by Eq.\ref{deftaud}. When MSD exceeds the 
threshold, $\sim 20$, signalled by the dashed line,  $\widehat C(t)$ drops below $0.1$. The threshold is virtually 
independent of the chain length and corresponds to about five bond lengths, i.e. about four monomer diameters.} 
\label{Msd(m)}
\end{figure}

\begin{figure}[t]
\begin{center}
\includegraphics[width=0.5\linewidth]{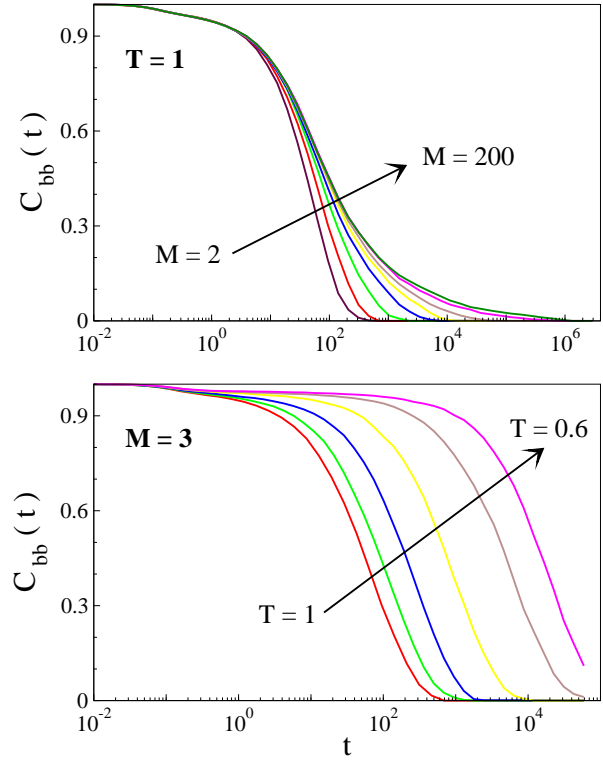} \\
\end{center}
\caption{Bond-bond correlation function $C_{bb}(t)$, Eq.\ref{bondbondcor}, for different chain lengths at $T=1$ 
(top) and for trimers at different temperatures (bottom). Color codes as in 
Fig.\ref{Msd(m)}. At long times $C_{bb}(t) \simeq \exp [-(t/\tau')^\beta]$ \cite{Molin09}. The stretching parameter ranges between $\beta_{M=200} \simeq 0.15$ and $\beta_{M=2} \simeq 0.9$.}
\label{deltaRb}
\end{figure}

\section{Results and discussion}
\label{resultsdiscussion}

\subsection{General aspects: monomer displacement and bond reorientation}
\label{MonMSD}

Fig.\ref{Msd(m)} shows an overview of the dependence of the monomer mean-square-displacement (MSD) on the chain length (top) and temperature (bottom). At short  
times ($t\sim 0.1$),  after the ballistic regime, the repeated collisions with the surroundings slow down the monomers 
and temporarily traps them in the cage formed by the first neighbours. The monomers escape from the cage on average 
within the time $\tau_\alpha$ (orange circles of Fig.\ref{Msd(m)}). We define, as in previous works
\cite{OurNatPhys,lepoJCP09,Puosi11,SpecialIssueJCP13,UnivSoftMatter11,UnivPhilMag11,OttochianLepoJNCS11,VoroBinarieJCP15,PuosiLepoJCPCor12,Puosi12,ElasticoEPJE15,Molin09,CapacciEtAl04},
the structural relaxation 
time by the equation $F_s(q_{max},\tau_\alpha)= \phi$ with $\phi = e^{-1}$ 
where $q_{max}$ is the maximum of the static structure factor and $F_s$ is the 
self-part of the intermediate scattering function \cite{HansenMcDonaldIIIEd}. 
It is worth noting that our polymer model complies with the temperature-time 
superposition principle, resulting in a {\it constant} (and moderate) 
stretching of $F_s(q_{max},t)$ \cite{OurNatPhys}. Then, even if other 
definitions of $\tau_\alpha$ are possible, e.g. by setting $\phi = 0.1$ 
\cite{Puosi11}, these alternatives differ from the present one of a {\it 
constant} factor of $\sim 3$, insignificant to the purposes of the present 
paper.  
In the present polymer model $\tau_\alpha$ is little dependent on $M$ since the chains are fully flexible \cite{Puosi11}. 

For times longer than $\tau_\alpha$ the polymer connectivity 
limits the monomer displacement and a series of different regimes are observed, 
all being characterized by subdiffusive motion, i.e. MSD $\propto t^\gamma$ 
with $\gamma < 1$ 
\cite{DoiEdwards,Rubinstein,Strobl,GoswamiPRE10,BennemannNat99}. The 
subdiffusive regimes are not of interest in the present paper. Their detailed 
description is found in textbooks \cite{DoiEdwards,Rubinstein,Strobl} and will 
be not repeated here for conciseness. Subdiffusive motion ends when the monomer 
moves a distance of the order of the chain size, the end-end mean square 
distance of the chain $R_{ee}^2$, and the diffusive regime is entered 
($\gamma=1$) \cite{DoiEdwards,Rubinstein,Strobl}. For the present polymer model 
one has $R_{ee}^2 = C_\infty (M-1) \; b^2$ with characteristic ratio $C_\infty 
\simeq 1.51$, in agreement with previous work  \cite{CapacciEtAl04}.

Let us define the time $\tau_D$ when MSD equals the end-end mean square distance of the chain $R_{ee}^2$:
\begin{equation}
\langle \Delta r^2(\tau_D )\rangle  = R^2_{ee}
\label{deftaud}
\end{equation}
$\tau_D \sim \tau_r$ where $\tau_r$ is the average reorientation time of the 
chain, i.e. the longest relaxation time of the correlation function of the 
end-end vector \cite{DoiEdwards}. For {\it entangled} polymers $\tau_r$ is also 
known as the disengagement time $\tau_d$ \cite{DoiEdwards,Rubinstein,Strobl}.
The onset time of the diffusive regime is strongly dependent on the chain 
length. In fact, for {\it unentangled} chains $\tau_D \sim \tau_r \propto M^2$, 
whereas for {\it entangled} chains $\tau_D \sim \tau_d \propto M^3$ 
\cite{DoiEdwards,Rubinstein,Strobl}. For $t \gtrsim \tau_{D}$, $\langle \Delta  
r^2(t)\rangle=6Dt$, where $D$ is the diffusion coefficient. From 
Eq.\ref{deftaud} one finds $D \sim R^2_{ee}/ 6 \tau_D$ so that, since $R_{ee}^2 
\sim M$, one recovers 
$D \propto M^{-1}$ for unentangled chains,  whereas $D \propto M^{-2}$ for entangled chains \cite{DoiEdwards,Rubinstein,Strobl}.

Another quantity of interest to the present study is the bond-bond correlation 
function: 
\begin{equation}
\label{bondbondcor}
C_{bb}(t) = \left \langle \frac{1}{(M-1) \; b^2} \sum_{i=1}^{M-1} {\bf b}_i(t) \cdot {\bf b}_i   \right \rangle_{N_c}
\end{equation}
Brackets denote the ensemble average over all the $N_c$ chains of the melt. Representative plots of $C_{bb}(t)$ are shown in Fig.\ref{deltaRb}. It is known that the bond-bond correlation function decays at long times as  a stretched exponential, $C_{bb}(t) \simeq \exp [-(t/\tau')^\beta]$  \cite{Molin09}.  The stretching parameter increases by decreasing the chain length. We find $\beta_{M=200} \simeq 0.15$ and $\beta_{M=2} \simeq 0.9$.

\begin{figure}[t]
\begin{center}
\includegraphics[width=0.5\linewidth]{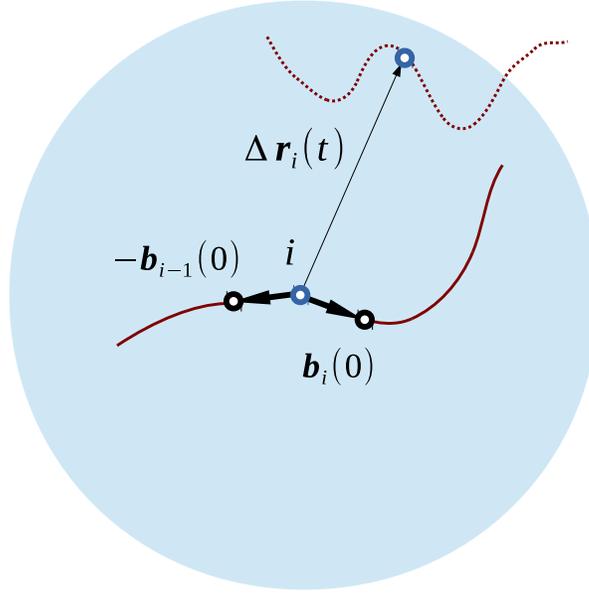}
\end{center}
\caption{Sketch of the i-th monomer displacement $\Delta \mathbf{r}_i(t)$ in a time $t$ and the initial orientations of 
the bonds with the adjacent monomers of a linear chain molecule. { Maximum correlation of $\Delta \mathbf{r}_i(t)$ with the initial orientations of the bonds is reached when the former is of the order of the bond length. At later times the correlation decay slowly and is not negligible if the monomer displaces by less than the radius of the colored sphere, about five times the bond length, see Sec.\ref{resultsCOA}.}}
\label{Cor}
\end{figure}

\subsection{Connectivity anisotropy (COA)}
\label{defcor}

\subsubsection{Definition}
\label{defCOA}
Let us consider a melt of $N_c$ linear polymer chains  with  $M$ monomers and $M-1$ bonds. The i-th monomer has position  $\mathbf{r}_i (t)$ at time $t$. It is linked  to the neighbouring monomers by bonds $\mathbf{b}_{i-1}(t)$ and $\mathbf{b}_{i}(t)$ with fixed length $b$, see Fig.\ref{Cor}, where:
\begin{equation}
\label{bond1}
\mathbf{b}_{i}(t) = \mathbf{r}_{i+1} (t)- \mathbf{r}_{i}(t) \qquad  1\leq i \leq M-1
\end{equation}
We consider the  monomer displacement in a time $t$, $ \Delta \mathbf r_i (t) 
\equiv \mathbf{r}_{i}(t) - \mathbf{r}_{i}(0)$, and its modulus $\left|  \Delta 
\mathbf r_i (t)  \right|$, see Fig.\ref{Cor}. We define the
following correlation function between the {\it direction} of the monomer displacement in a time $t$ and the initial orientation of the bonds linking the monomer to the adjacent ones ($\mathbf{b}_{i} \equiv \mathbf{b}_{i}(0)  $):
\begin{equation}
\widehat C(t) = \left \langle \frac{1}{2} \frac{1}{M-1}  \sum_{i=1}^{M}  
\frac{\Delta \mathbf r_i (t)}{ \left|  \Delta \mathbf r_i (t)  \right| } \cdot \frac{({\bf b}_i - {\bf b}_{i-1})}{b} \right \rangle_{N_c}
\label{tetabond}
\end{equation}
It is understood that ${\bf b}_M = {\bf b}_0 = 0$. The definition is independent of the choice of the end monomer labelled as $i=1$. The sum in Eq.\ref{tetabond}  is divided by twice the total number of bonds per chain. Henceforth, $\widehat C(t)$ will be referred to as the { connectivity } anisotropy function.

$\widehat C(t)$ vanishes at both short and long times. In fact, at short times the monomer displacement is ballistic, whereas at long times it is diffusive. In both regimes  the displacement direction is not correlated with the initial bond arrangement. To better appreciate the major features of $\widehat C(t)$, it is worthwhile to define the companion function:
\begin{equation}
\label{tetaB_app}
\widetilde C(t) = \frac{ b }{2 \, \langle \left|  \Delta \mathbf r_m (t)  \right|  \rangle} \, \left [ 1 - C_{bb}(t) \right ] 
\label{tetaB_appn}
\end{equation}
$\widetilde C(t)$ is derived in \ref{Appendix}. It is an effective approximation of  the { connectivity } anisotropy function $\widehat C(t)$, see Sec.\ref{resultsCOA}. Since $C_{bb}(t)$ vanishes at long times, see Fig.\ref{deltaRb}, $\widetilde C(t)$ decays as:
\begin{equation}
\widetilde C(t) \simeq \frac{ b }{2 \; \langle \left|  \Delta \mathbf r_m (t)  \right|  \rangle} 
\label{ctildeAppr}
\end{equation}
At long times in the diffusive regime, $ \left|  \Delta \mathbf r_m (t)  \right| \simeq \langle \Delta r^2( t )\rangle^{1/2} \simeq \alpha \, t^{1/2}$ with  $\alpha = R_{ee} \tau_D^{-1/2}$ from Eq.\ref{deftaud}. Then,  Eq.\ref{ctildeAppr} yields:
\begin{equation}
\widetilde C(t) \simeq \frac{b }{2 \; R_{ee}}  \; \left ( \frac{\tau_D}{t} \right )^{1/2}  \hspace { 6mm} t \gg \tau_D
\label{ctildeAppr2}
\end{equation}
Eq.\ref{ctildeAppr2} emphasizes the slow decay of the { connectivity } anisotropy function.

The derivation of $\widetilde C(t)$ does not make any assumption on the chain length. However, to provide more insight into our results, it proves useful to derive in  \ref{Appendix2}  the short-chain limit of $\widetilde C(t)$. To this aim, we neglect the role of the entanglements and resort to the Rouse gaussian theory of polymer dynamics which pictures the chains as "phantoms", i.e. perfectly crossable, and dissolved in a structureless environment \cite{DoiEdwards,Rubinstein,Strobl,Molin09}. The short-chain limit of $\widetilde C(t)$ will be denoted as $\widetilde C^R(t)$.

\begin{figure}[t]
\begin{center}
\includegraphics[width=0.5 \linewidth]{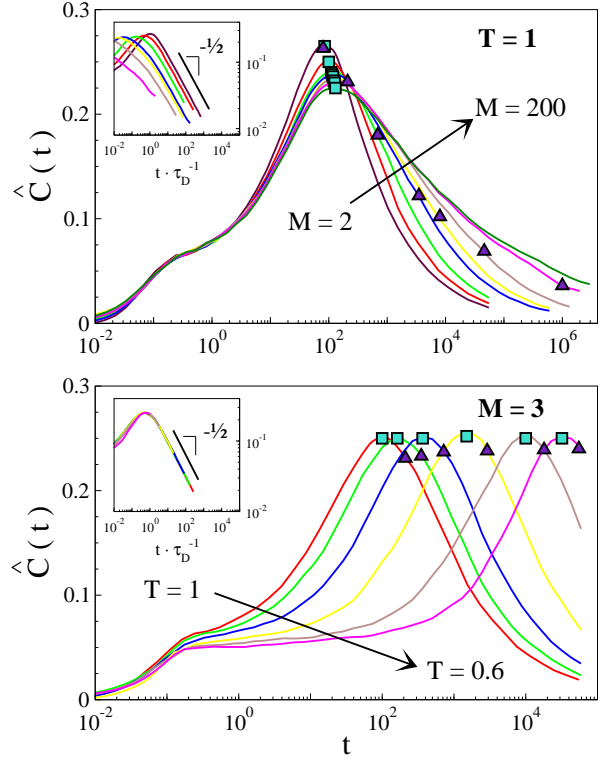}
\end{center}
\caption{ { Connectivity } anisotropy function $\widehat C(t)$, Eq.\ref{tetabond}. Top: dependence on the chain length for $T=1$. Bottom: 
dependence on the temperature for trimers ($M=3$).  Symbols and color codes as in Fig.\ref{Msd(m)}. The insets of the 
two panels focus on the long-time part confirming the asymptotic decay predicted by Eq.\ref{ctildeAppr2}. $\widehat C(t)$ reaches the maximum at $\tau_{max}$ (cyan squares) when the 
monomer displaces by one diameter on average (see Fig.\ref{Msd(m)}). Note that the position and the height of the 
maximum are nearly independent of the chain length. The maximum height and the residual correlations at the onset of the 
diffusive regime at  $\tau_D$ (violet triangles) are nearly independent of the temperature.}
\label{tetab}
\end{figure}

\subsubsection{Results}
\label{resultsCOA}

Fig.\ref{tetab} plots the { connectivity } anisotropy 
function $\widehat C(t)$, Eq.\ref{tetabond}. It is seen that $\widehat C(t)$ vanishes at short times, owing to the 
initial missing correlations between the directions of the displacement and the bond orientation. { Only a small bump is observed at $t \sim 0.175$, the average time needed to reverse the particle velocity by collisions with the cage of the first neighbours (see Sec.\ref{resultsLOA} and Fig.\ref{test8}), signalling the weak influence of the local order on $\widehat C(t)$. Later, a peak is observed at $\tau_{max}$}, where $\widehat C(t)$ is maximum. 
Both the  position and the height of the peak are nearly independent of the chain length, see Fig.\ref{tetab} (top). This is 
understood by noting that the monomer displaces only by about one bond length on average at $\tau_{max}$, see 
Fig.\ref{Msd(m)}, and, therefore, does not experience the constraints posed by all the connected structure. At longer 
times $\widehat C(t)$ decreases, more slowly for longer chains, and finally vanishes as $\widehat C(t) \sim t^{-1/2}$ 
for  $t \gg \tau_D$, see Fig.\ref{tetab} (insets), in agreement with the long-time decay predicted by Eq.\ref{ctildeAppr2}. 
Note that the maximum height and the residual correlations at 
the onset of the diffusive regime at  $\tau_D$ are nearly independent of the temperature, see Fig.\ref{tetab} (bottom). 

It is customarily said that the correlations in a polymer melt vanish when the chain moves it own size $\sim R_{ee}$.  This statement is scrutinized in Fig.\ref{tetab} where the violet triangles  mark the residual  anisotropy at time $\tau_D$, i.e. the time needed to displace a monomer of $R_{ee}$, see Eq.\ref{deftaud}. It is seen that $\widehat C(\tau_D)$  is not negligible for $M \lesssim 15$ since it exceeds $0.1$. In particular, $\widehat C(\tau_D) \sim 0.12$ for a decamer. 
This residual correlation is captured by $\widetilde C(\tau_D)$, Eq.\ref{ctildeAppr2}. In fact, by reminding the chain-length dependence of the end-end distance, see Sec.\ref{MonMSD}, one finds $\widetilde C(\tau_D) \sim  b/ (2 R_{ee}) = 1/\sqrt{4 C_\infty (M-1)}  \sim 0.14 $ for $M=10$, in good agreement with $\widehat C(\tau_D)$.

\begin{figure}[t]
\begin{center}
\includegraphics[width=  0.5\linewidth]{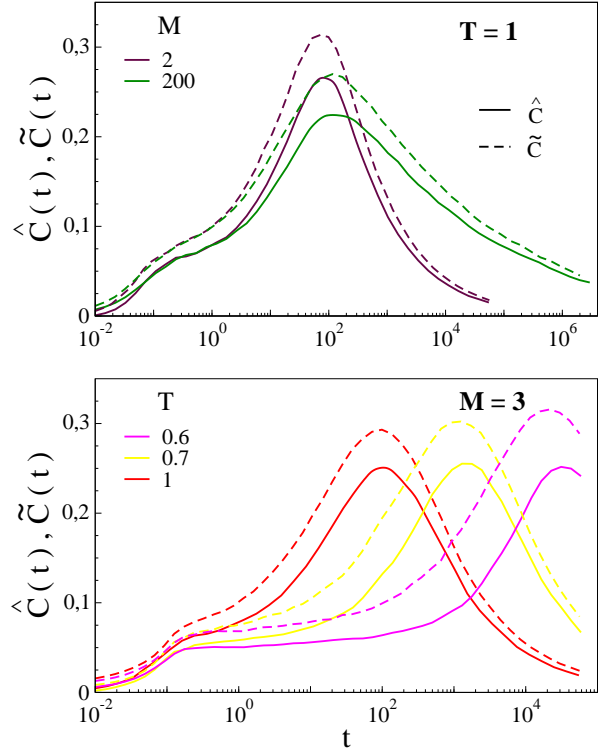} 
\end{center}
\caption{Comparison of the anisotropy function $\widehat C(t)$, Eq.\ref{tetabond}, with the first-order approximation,  
$\widetilde C(t)$, Eq.\ref{tetaB_appn}, for selected chain lengths  (top) and temperatures (bottom).}
\label{approx}
\end{figure}

The good agreement at $\tau_D$ prompts us to compare at any time the exact  { connectivity } anisotropy function $\widehat C(t)$, Eq.\ref{tetabond}, and the first-order approximation  $\widetilde C(t)$, Eq.\ref{tetaB_appn}. The results are shown in Fig.\ref{approx}. It is seen that the agreement is satisfactory with average deviations of about 
15 $\%$. 

Eq.\ref{ctildeAppr} suggests that the anisotropy of the random walk, after the maximum at $\tau_{max}$, decays at $0.1$ when the monomer 
displaces on average by {\it five} bond lengths, corresponding to about {\it four} particle diameters, independently of the chain length. The prediction is satisfactory. Indeed, by inspecting Fig.\ref{Msd(m)} and Fig.\ref{tetab}, one finds that $\widehat C(t) = 0.1$ if  the root mean square displacement  $\simeq 4.61$ bond lengths, for {all} the chain lengths under study. Fig.\ref{Cor} provides a 
pictorial representation of the situation. The finding may be also expressed in terms of the Kuhn length $\ell_K$,  the local stiffness of the polymer chain, to get rid of the details of the MD model and resort to a quantity experimentally accessible. It is found \cite{Rubinstein}:  
\begin{eqnarray*}
\ell_K  &\equiv& \frac{R^2_{ee}}{(M-1) b} \\
&=& C_\infty \, b \\
&=& 1.51 \; b
\end{eqnarray*}
One concludes that the anisotropy of the random walk { due to the connectivity} is not negligible over monomer displacements as large as about {\it three} Kuhn segments. 
Notice that two sites of the same chain have correlated geometry if they are spaced by about {\it one} Kuhn length. 
This means that the spatial decay of the anisotropy is not driven by the local stiffness of the chain but it follows 
from the slow spatial decay of the correlations, Eq.\ref{ctildeAppr}.

\begin{figure}[t]
\begin{center}
\includegraphics[width= 0.5 \linewidth]{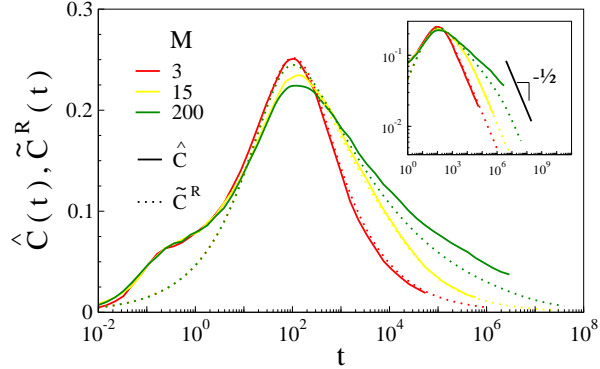} 
\end{center}
    \caption{{ Comparison of the connectivity anisotropy function, $\widehat 
C(t)$, with the short-time limit ${\widetilde C}^R(t)$, Eq.\ref{CrHat} with { 
$\kappa = 0.725$}, for selected chain lengths and $T=1$}. Color codes as in 
Fig.\ref{Msd(m)}. The inset focus on the long-time decay. The characteristic 
time $\tau$ of Eq.\ref{masdR} and Eq.\ref{eq:selfRouseprediction} has been set 
to match the peak position of $\widehat C(t)$, whereas the parameter { 
$\kappa$} has been adjusted to best-fit Eq.\ref{CrHat} to the results for 
$M=3$. The values of both $\tau$ and $\kappa$ are the same for $M=3,15,200$. 
Note that ${\widetilde C}^R(t)$ agrees very nicely with $\widehat C(t)$ for $t 
\gtrsim 10$ and $M=3,15$. It underestimates the connectivity anisotropy at 
short times and,  for $M=200$, at long times.}
\label{rouse}
\end{figure}

{ Fig.\ref{rouse} compares the { connectivity } anisotropy function, $\widehat C(t)$, with the short-chain limit $\widetilde C^R(t)$, derived in \ref{Appendix2}. $\widetilde C^R(t)$ is smaller at short times, $t \lesssim 10$. This is readily interpreted by reminding that  $\widetilde C^R(t)$ is derived in terms of the Rouse theory, which pictures the chain surroundings as structureless \cite{DoiEdwards,Molin09}, thus it misses the enhancement of the anisotropy due to local order, see Sec.\ref{appl}. The agreement between $\widetilde C^R(t)$ and  the { connectivity } anisotropy function $\widehat C(t)$ becomes extremely good for $t \gtrsim 10$ and $M=3,15$. For $M=200$, $\widetilde C^R(t)$ underestimates $\widehat C(t)$ after the maximum. This is clear evidence that the COA loss  is slowed down by the presence of chain entanglements which are neglected by $\widetilde C^R(t)$, relying on the Rouse theory which pictures the chains as  "phantoms", i.e. perfectly crossable \cite{DoiEdwards,Molin09}. For the present models entanglements play a significant role for $M \gtrsim 80$, see Sec.\ref{numerical}.
}

\begin{figure}[t]
\begin{center}
\includegraphics[width= 0.33 \linewidth]{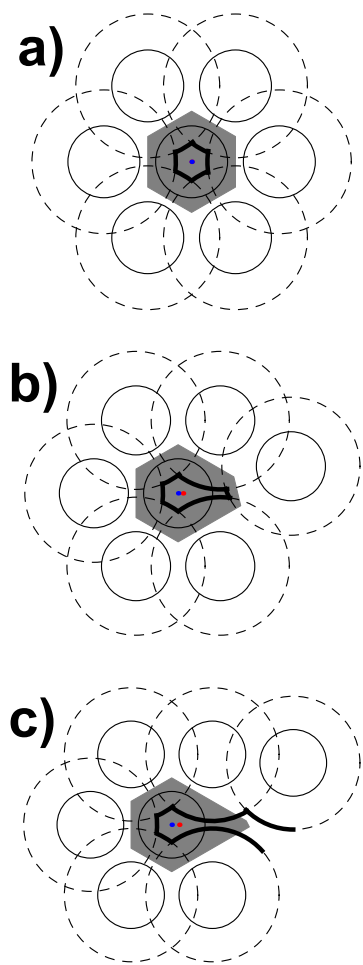} 
\end{center}
\caption{Schematic 2D view of a particle trapped by the cage of six first 
neighbours. Particles are sketched as hard disks with continuous-lines. The 
Voronoi cell is the grey area. Note that the VP vertices are close to the voids 
between the particles. For a given - fixed - configuration of the first shell, 
the center of the trapped particle may span a well-defined free-volume 
represented by the region with thick borders. The region is built by noting 
that the centres of two hard disks  cannot be closer than their diameter $d$. 
The dashed circles have radius $d$ and are the boundaries of the excluded 
regions to the central particle for each of the six neighbours. The three 
panels differ by different positions of the right-most particle. They refer to 
a ideally-ordered packed (top), deformed but still close (middle) and open 
(bottom) cage configurations. The blue dot is the position of the trapped 
particle which, together with the position of the surrounding particles, set 
the Voronoi cell with centroid of the vertices located at the red dot. The LOA 
axis is the line joining the blue dot with the red one (not drawn for clarity 
reasons). It provides a good indication of the  {\it direction} along which the 
VP and the free-volume are deformed. Starting from a given seven-particle 
configuration, at later times the trapped particle  tends to  move initially 
along the LOA axis  \cite{Rahman66,DouglasAspehricity}. The LOA axis is not 
defined in the extreme, and very rare \cite{LocalOrderJCP13}, case of 
ideally-ordered cage.}
\label{VoronoiFreeVolume}
\end{figure}

\begin{figure}[t]
\begin{center}
\includegraphics[width= 0.5 \linewidth]{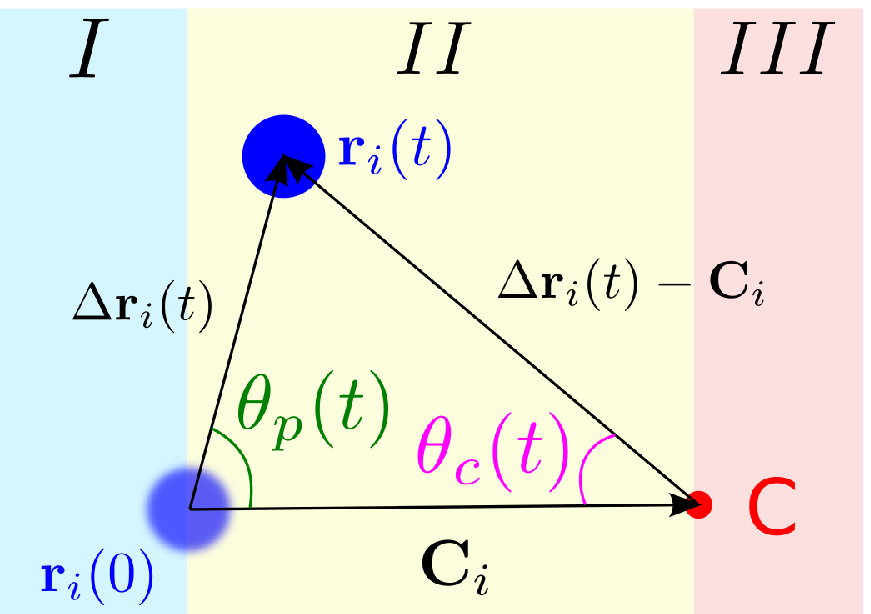} 
\end{center}
\caption{Quantities of interest to characterize the local-order anisotropies $\langle\cos\theta_p(t)\rangle$ and 
$\langle\cos\theta_c(t)\rangle$. Particle position at time $0$ and $t$ is marked by blue dots and the centroid of the 
VP vertices, Eq.\ref{centroid}, at the initial time is marked by red dot. The highlighted regions are limited by 
planes perpendicular to  $\mathbf C_i$  and  passing through either the initial position of the particle or the 
centroid.}
\label{disegnino}
\end{figure}

\subsection{Local-order anisotropy (LOA)}
\label{appl}

Sec.\ref{defcor} discussed the COA anisotropy of the monomer random-walk which is maximum when the displacement is about the bond length. For smaller displacements the local order  plays a major role to drive the anisotropy of the monomer displacement.  The present Section investigates and discusses this aspect.

\subsubsection{General aspects}
\label{locOrdGenAspect}
LOA anisotropy is characterised in the present work by employing the vertices of the VP cell. That approach is well-known for atomic liquids  \cite{Rahman66} and granular matter \cite{DouglasAspehricity} but, as far as we know, never applied to molecular liquid with competing COA anisotropy. The main motivation to concentrate on the VP vertices is that they are close to the voids between the particles and thus signal the weak spots of the cage, see Fig.\ref{VoronoiFreeVolume}. To expose the correlation between the local order and the monomer dynamics, we consider the available free-volume. 
Rigorously, the geometrical free volume is defined as the volume over which the centre of the trapped sphere can translate, being fixed the cage configuration \cite{SastryEtAl98}. 
The VP cell and the free-volume region are, in general, not coinciding even if they exhibit some qualitative resemblance, e.g. see the examples of well-packed, Fig.\ref{VoronoiFreeVolume}(a),   mildly, Fig.\ref{VoronoiFreeVolume}(b), and heavily distorted, Fig.\ref{VoronoiFreeVolume}(c), cages. Cages are usually quite distorted in both atomic and molecular liquids \cite{LocalOrderJCP13}. 
The distortion is due to defective packing leading to the opening of one weak 
spot of the cage, or more. As an example, Fig.\ref{VoronoiFreeVolume} sketches 
three snapshots of an elementary opening process due to the rearrangement of 
one single member of the first shell. It is seen that the VP vertices being 
closer to the rearranging member approach the widening weak spot. The centroid, 
the center of mass of the VP vertices, does the same. 
Fig.\ref{VoronoiFreeVolume} shows that, for a given distorted cage, the 
free-volume is distorted too and the elongation of  the free-volume occurs 
nearly along the same axis of the VP cell. This axis is approximated by the LOA 
axis, the line joining the centroid with the centre of the trapped monomer 
\cite{Rahman66,DouglasAspehricity}.  The importance of the LOA axis resides in 
the previous finding \cite{Rahman66,DouglasAspehricity}, which may be hinted at 
in Fig.\ref{VoronoiFreeVolume} and will be  substantiated in 
Sec.\ref{resultsLOA} for a molecular liquid as well, that, {\it starting from a 
given configuration at $t_0$, at later times the trapped particle  tends to  
move initially along the LOA axis set at $t_0$}.

\subsubsection{Definition}
\label{locOrdDef}

On the basis of the discussion in Sec.\ref{locOrdGenAspect}, we define LOA axis 
at the initial time as \cite{Rahman66,DouglasAspehricity}:
\begin{equation}\label{uC}
\hat{\bf u}^C_i = \frac{\mathbf C_i}{ |\mathbf C_i|} 
\end{equation} 
$\mathbf C_i$ is the position of the centroid, the center of mass of the VP vertices, with respect to the position of 
the $i$-th particle at the initial time, see Fig.\ref{disegnino}:
\begin{equation}
 \mathbf C_i = \frac{1}{N_{v, \, i}}\sum_{j=1}^{N_{v, \, i}} \mathbf v^j_i
 \label{centroid}
\end{equation}
where $N_{v, \, i}$ and  $\mathbf v^j_i$ are the number of vertices and  the position of the VP $j$-th vertex with 
respect to the position of the $i$-th particle at the initial time, respectively. 

In order to investigate the LOA anisotropy of the i-th particle we investigate 
two distinct order parameters. First, following previous studies 
\cite{Rahman66,DouglasAspehricity}, we consider the correlation between the LOA 
axis set at the initial time and the direction of the displacement of the 
$i$-th monomer in a time lapse $t$, $ \hat{\bf u}_i (t) \equiv \Delta\mathbf 
r_i(t)/ \left | \Delta\mathbf r_i(t) \right |$, see Fig.\ref{disegnino}:
\begin{equation}
\label{cosTP}
 \langle\cos\theta_p(t)\rangle = \frac{1}{N} \sum_{i=1}^N \hat{\bf u}_i (t)  \cdot \hat{\bf u}^C_i
\end{equation}
Furthermore, we also consider the correlation between the LOA axis set at the initial time and  the direction of the particle position at time $t$ with respect to the centroid position at the initial time, $\big[\Delta\mathbf r_i(t) -  \mathbf C_i \big] 
/ \big|\Delta\mathbf r_i(t) -  \mathbf C_i \big|$, see Fig.\ref{disegnino}:
\begin{equation} 
\label{cosTC}
 \langle\cos\theta_c(t)\rangle = \frac{1}{N} \sum_{i=1}^N \frac{\big[\Delta\mathbf r_i(t) -  \mathbf C_i \big] 
}{\big|\Delta\mathbf r_i(t) -  \mathbf C_i \big|} \cdot 
\big[-\hat{\bf u}^C_i \big]
\end{equation}
Complete isotropy yields $\langle\cos\theta_i\rangle = 0$ ($i=p,c$). Perfect alignment of $ \hat{\bf u}_i (t)$ 
with respect to $\hat{\bf u}^C_i$ yields $ \langle\cos\theta_p\rangle = 1$ whereas perfect alignment of $\big[\Delta\mathbf 
r_i(t) -  \mathbf C_i \big]$ with respect to $- \hat{\bf u}^C_i$ yields $ \langle\cos\theta_c\rangle = 1$. Furthermore, 
if the monomer displacement is large with respect to $ |\mathbf C_i|$, $\theta_p(t) \simeq \pi - \theta_c(t)$ and  
$\langle\cos\theta_p(t)\rangle \approx -\langle\cos\theta_c(t)\rangle$.  The order parameters defined by Eq.\ref{cosTP} 
and Eq.\ref{cosTC} provide complementary information. By referring to Fig.\ref{disegnino}, positive values of  
$\langle\cos\theta_c(t)\rangle$ signal that the particle is preferentially located in regions I and II, whereas positive 
values of $\langle\cos\theta_p(t)\rangle$ denote preferential location of the 
particle in regions II and III. Fig.\ref{VoronoiFreeVolume} suggests that 
regions I and III are populated, initially,  by monomers well inside the cage 
and close to the weak spots of the cage, respectively, being region II a 
transition zone.

It must be pointed out that the anisotropies $\langle\cos\theta_p(t)\rangle$ and $\langle\cos\theta_c(t)\rangle$ 
are not restricted to polymer systems, but are well-defined for atomic liquids too.

\begin{figure}[t!]
\begin{center}
\includegraphics[width=  0.5 \linewidth]{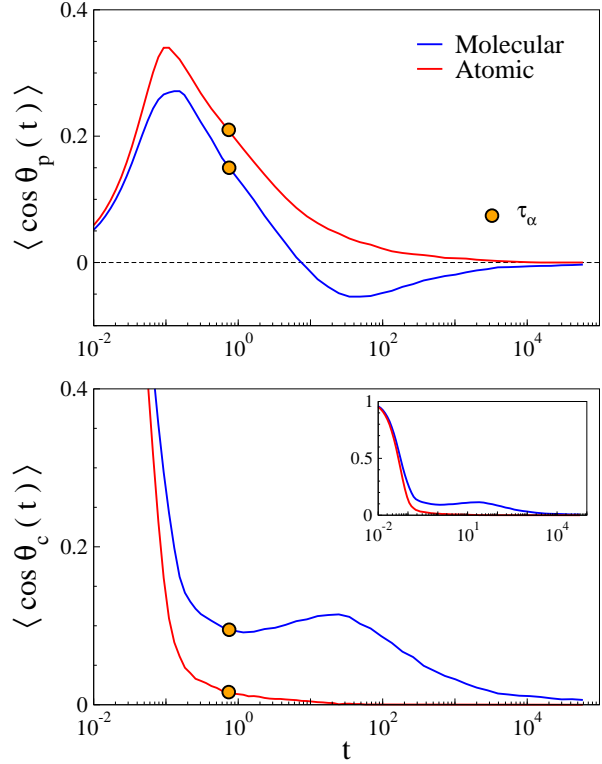} 
\end{center}
\caption{Comparison between the local-order anisotropies $\langle\cos\theta_p(t)\rangle$  (top) and 
$\langle\cos\theta_c(t)\rangle$  (bottom) in a melt of trimers (blue) and an atomic liquid (red) with the same temperature ($T=1.5$) and density. The inset in the bottom panel shows the complete decay of $\langle\cos\theta_c(t)\rangle$.}
\label{test7}
\end{figure}

\begin{figure}[t!]
\begin{center}
\includegraphics[width=  0.5 \linewidth]{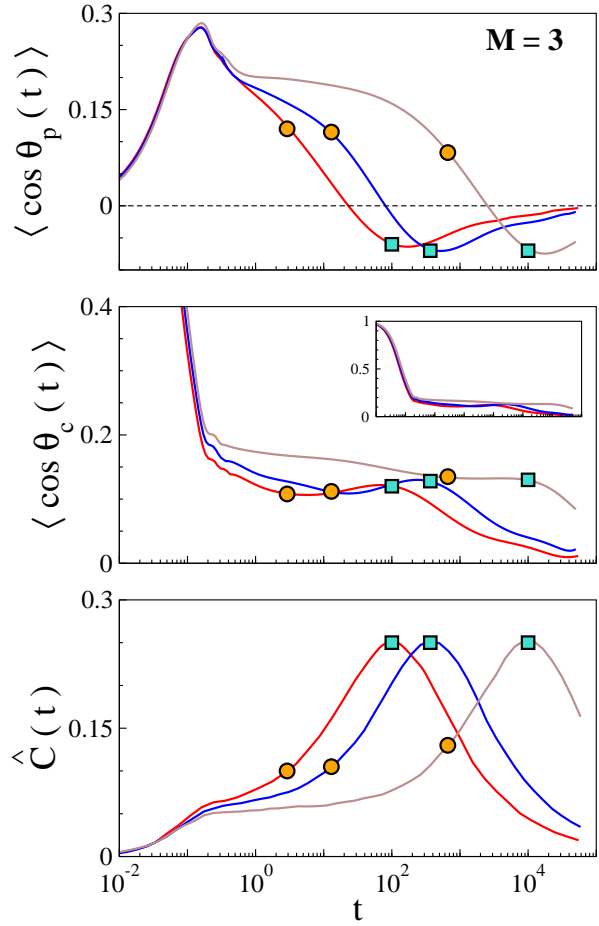} 
\end{center}
\caption{Local-order anisotropies , $\langle\cos\theta_p(t)\rangle$ (top) and 
$\langle\cos\theta_c(t)\rangle$ (middle) for trimers at different temperatures. The inset of the middle panel shows the full decay of $\langle\cos\theta_c(t)\rangle$. For comparison, the corresponding connectivity anisotropy $\widehat C(t)$ is plotted in the bottom panel. Symbols and color codes as in Fig.\ref{Msd(m)}.
}
\label{test8}
\end{figure}

\subsubsection{Results}
\label{resultsLOA}
To provide comparison with, and extend, previous studies about $\langle\cos\theta_p(t)\rangle$ in atomic liquids \cite{Rahman66},  Fig.\ref{test7} compares the LOA anisotropies of the melt of trimers under study and an atomic liquid with the same temperature and density. 
Let us start just with $\langle\cos\theta_p(t)\rangle$, Eq.\ref{cosTP}. It is plotted in Fig.\ref{test7} (top). At very 
short times the direction of the particle displacement $\hat{\bf u}_i (t)$  is almost isotropic and 
$\langle\cos\theta_p(t)\rangle$ is small. Referring to 
Fig.\ref{VoronoiFreeVolume}, this corresponds to the early stage of the cage 
exploration by the trapped monomer. Later, $\langle\cos\theta_p(t)\rangle$ 
increases and reaches the maximum at $t\sim0.175$. One notices that, for both 
the molecular and the atomic liquids, the maximum corresponds to the minimum of 
the velocity correlation function (not shown), i.e. the time  needed by most 
particles to reverse their initial velocity due to the collision with the cage 
of the first neighbours. The presence of a well-defined maximum of 
$\langle\cos\theta_p(t)\rangle$ confirms also for a molecular liquid the 
initial tendency of the trapped particle to move along the LOA axis already 
evidenced in atomic \cite{Rahman66} and granular matter 
\cite{DouglasAspehricity}. It is clear indication that {\it initially} there is 
correlation between the  local structure and the particle displacement. The 
reduction of the maximum of $\langle\cos\theta_p(t)\rangle$ from the ideal unit 
value is manifestation of the partial misalignment of the monomer displacement 
with respect to the LOA axis. Interestingly, the connectivity acts as a 
constraint in the rattling motion of the monomer inside the cage, due to the 
bonds linking the trapped monomer to one or two of the closest monomers.  This 
results in 
additional misalignment, thus leading to the decrease of the maximum of $\langle\cos\theta_p(t)\rangle$ of $\sim 20 \%$  with respect to the atomic liquid. After the maximum the anisotropy $\langle\cos\theta_p(t)\rangle$ decreases and  vanishes in a monotonous way in the atomic liquid. Differently, in the liquid of trimers a minimum is observed. This regime, which is controlled by the connectivity, will be analysed in detail below.
We now discuss $\langle\cos\theta_c(t)\rangle$, Eq.\ref{cosTC} plotted in Fig.\ref{test7} (bottom). At very short times the displacement $\Delta\mathbf r_i(t)$ is small and $\langle\cos\theta_c(t)\rangle \sim 1$. Then,  the anisotropy drops up to $t\sim0.175$ where a knee is observed. The knee corresponds to the maximum of $\langle\cos\theta_p(t)\rangle$. For longer times $\langle\cos\theta_c(t)\rangle$ vanishes at about $\tau_\alpha$ in the atomic liquid whereas it persists at much longer times in the liquid of trimers. This regime, which is controlled by the connectivity, will be analysed in detail below. 
All in all, Fig.\ref{test7} suggests that, even for times shorter than the structural relaxation time  the influence of the local order on the monomer displacement is partially decreased by the connectivity, at longer times the latter favours the persistence of the local order. 

Fig.\ref{test8}  examines in detail the LOA anisotropies of the liquid of trimers. Fig.\ref{test8} (top) plots $\langle\cos\theta_p(t)\rangle$ at different temperatures. It is seen that the region around the maximum is virtually unaffected by the temperature changes, whereas for $t \gtrsim 0.7$ a significant slowing-down is observed by decreasing the temperature. A similar behaviour is observed in $\langle\cos\theta_c(t)\rangle$, see  Fig.\ref{test8} (middle) even if the influence of the temperature appears earlier in time since it is already visible after the knee at $t\sim0.175$. At times longer than the structural relaxation time a minimum of $\langle\cos\theta_p(t)\rangle$ is observed in coincidence with the local maximum of $\langle\cos\theta_c(t)\rangle$. The coincidence follows by the approximate relation $\langle\cos\theta_p(t)\rangle \approx -\langle\cos\theta_c(t)\rangle$, which holds at long times, see Sec.\ref{locOrdDef}.  By comparison with the connectivity anisotropy function $\widehat C(t)$, Fig.\ref{test8} (bottom) one realises that the minimum of $\langle\cos\theta_p(t)\rangle$  and the local maximum of $\langle\cos\theta_c(t)\rangle$ just mirror the region of the COA maximum which occur for monomer displacements of about one bond length, see Fig.\ref{Msd(m)}.

Fig.\ref{test7} and Fig.\ref{test8} encompass our results about LOA and COA. At short times, before the structural relaxation, connectivity and local order  play different roles. In fact, the connectivity is antagonistic and decreases LOA, Fig.\ref{test7} (top), viceversa, COA is slightly enhanced by the local order, Fig.\ref{test8} (bottom). After structural relaxation, the connectivity ensures the persistence of LOA up to displacements as large as about one bond length, or one particle diameter, Fig.\ref{test7} (bottom) and Fig.\ref{test8} (middle). 

It is interesting to interpret the localization of the monomer in the three regions defined in Fig.\ref{disegnino} in the light of our results on LOA anisotropy. At very short times the monomer is mainly in region I and, sligthly more, in region II ( $\langle\cos\theta_p(t)\rangle$ and $\langle\cos\theta_c(t)\rangle$ both positive). Then, the monomer approaches the centroid and region II and region III become more populated. As a result, $\langle\cos\theta_p(t)\rangle$ increases up to the maximum and  $\langle\cos\theta_c(t)\rangle$ decreases. After the structural relaxation $t \gtrsim \tau_\alpha$, the direction of the monomer displacement with respect to both the original position and the centroid, randomises in the {\it atomic} liquid, yielding the monotonous loss of LOA anisotropy, see Fig.\ref{test7}. Instead, in the {\it molecular} liquid where COA is present,  both the change of sign of $\langle\cos\theta_p(t)\rangle$ and the increase of $\langle\cos\theta_c(t)\rangle$  following the structural relaxation, see Fig.\ref{test7} and Fig.\ref{test8}, suggest that the bonds resist the randomization of the monomer displacement and tend to pull back the monomer, thus enriching the monomer population in region I.  The strength of the effect is driven by COA since the latter is a measure of the correlation between the monomer displacement and the arrangements of the bonds tethering the monomer to the adjacent ones, see Sec.\ref{defCOA}.

The above results about LOA provide insight into our previous finding that the 
rattling amplitude of the monomer in the cage of the closest neighbours has 
poor correlation with the cage shape in the same liquid of trimers under study 
here \cite{VoroBinarieJCP15,VoronoiBarcellonaJNCS14}. The mean square rattling 
amplitude was evaluated as the MSD at $t \simeq 1$, a quantity exhibiting 
universal correlation with the structural relaxation  
\cite{OurNatPhys,lepoJCP09,Puosi11,SpecialIssueJCP13,UnivSoftMatter11,DouglasCiceroneSoftMatter12,DouglasStarrPNAS2015,UnivPhilMag11,OttochianLepoJNCS11,CommentSoftMatter13}. We now see in Fig.\ref{test8} that the LOA anisotropies, $\langle\cos\theta_p(t)\rangle$ and $\langle\cos\theta_c(t)\rangle$  are rather small at $t \simeq 1$. 
The small LOA value strongly suggests that the sole consideration of the local scale is not enough to understand the microscopic origin of the rattling amplitude of the trapped monomer in the cage, in agreement with previous conclusions pinpointing the major role of collective effects on larger length scales \cite{PuosiLepoJCPCor12,PuosiLepoJCPCor12_Erratum,BerthierJackPRE07}.

The consideration of linear chains longer than trimers changes only qualitatively the conclusions reached up to now about the interplay between LOA and COA. For conciseness, they are only briefly summarized. 
For times shorter than $\tau_\alpha$ the key parameter is the average {\it effective}  number of bonds per monomer $2(1 - 1/M)$ \cite{LocalOrderJCP13}, so that the effective bond constraint on the rattling motion in the cage increases with the chain length. As a result, the maximum  of $\langle\cos\theta_p(t)\rangle$ decreases by increasing the chain length, a finding that is readily interpreted by reminding that the connectivity disturbs the alignment of the monomer displacement with the LOA axis, see Fig.\ref{test7} (top). For times longer than than $\tau_\alpha$ we know that the COA maximum increases with the chain length and the decay is slowed down, see Fig.\ref{tetab}. As a consequence, being the LOA anisotropies driven by COA at intermediate and long times, the local maximum of $\langle\cos\theta_c(t)\rangle$ increases, the local minimum of $\langle\cos\theta_p(t)\rangle$ decreases and their decay at long times slows down.

\section{Conclusions}
\label{concl} 

Extensive MD simulations on melts of linear fully-flexible chains ranging from dimers up to  entangled polymers are performed to investigate the anisotropy of the monomer random walk. We consider  the roles of both the local geometry and the connectivity.

We first scrutinise the connectivity anisotropy (COA), the correlation between the initial bond orientation and the subsequent direction of the monomer displacement. At times comparable with the time needed by most particles to reverse their initial velocity due to the collision with the cage of the first neighbours, local order tends to {\it enhance} slightly COA. Later, we find  the peculiar non-monotonous  time-dependence of COA, peaking when the monomer displacement is of the order of the bond length and vanishing as $t^{-1/2}$ at longer times. COA is slowed down by the chain entanglements and  becomes negligible when  the displacements is as large as about five bond lengths, i.e. about four monomer diameters or three Kuhn lengths.

As to the local-order anisotropy (LOA), attention is paid to the LOA axis, the line joining at a given time $t_0$ the positions of the monomer trapped in the cage of the first neighbours and the related VP centroid, providing indications on the {\it initial} direction of the monomer displacement at times later than $t_0$. We consider two complementary LOA metrics by correlating  the LOA axis at $t_0$ with the subsequent time evolution of the directions of either the monomer displacement or the particle position with respect to the initial centroid position.
We conclude, also by comparison with a reference atomic liquid, that at times shorter  than $\tau_\alpha$ LOA  is affected by the shape of the cage of the closest neighbours with {\it antagonistic} effect by the connectivity. Differently, 
after structural relaxation, the connectivity {\it favours} the persistence of LOA up to displacements as large as about one bond length, or one particle diameter. 

Our results strongly suggest that the sole consideration of the local order is not enough to understand the microscopic origin of the rattling amplitude of the trapped monomer in the cage of the neighbours. 

\ack
A generous grant of computing time from IT Center, University of Pisa and ${}^{\circledR}$Dell Italia is gratefully acknowledged.

\appendix
\section{Derivation of the function $\widetilde C(t)$, Eq. \ref{tetaB_appn}}
\label{Appendix}
The function $\widetilde C(t)$, Eq. \ref{tetaB_appn}, is an approximation of the connectivity anisotropy function $\widehat C(t)$, Eq.\ref{tetabond}. The latter has the form $\widehat C(t) = \langle X_m/Y_m \rangle $ where $X_m$ and $Y_m$ are quantities referred to the $m$-th monomer:
\begin{eqnarray}
X_m  &=&  \Delta \mathbf r_m (t) \cdot \frac{({\bf b}_m - {\bf b}_{m-1})}{b} \frac{M}{2(M-1)} \\
Y_m  &=& \left|  \Delta \mathbf r_m (t)  \right| 
\end{eqnarray}
and the brackets denotes the average over all the {\it monomers},  that is 
\begin{equation}
\left \langle Z_m \right \rangle = \left \langle \frac{1}{M} \sum_{i=1}^{M}  Z_i \right \rangle_{N_c}
\label{average}
\end{equation}
One evaluates the ratio $X_m/ Y_m$ by expanding the random variables $X_m$ and $Y_m$ around their averages \cite{RatioDistr06}.  At first order, after suitable average, one has $\widehat C(t) = \langle X_m  /  Y_m \rangle \simeq \langle X_m \rangle / \langle Y_m \rangle \equiv \widetilde C(t)$ with :
\begin{eqnarray}
\widetilde C(t) = \frac{ b }{\langle \left|  \Delta \mathbf r_m (t)  \right|  \rangle} \, C(t)
\label{tetaB_appnAPP}
\end{eqnarray}
where
\begin{equation}
\label{bonddisplcor}
C(t) = \left \langle \frac{1}{2} \frac{1}{M-1} \frac{1}{b^2}  \sum_{i=1}^{M}  \Delta \mathbf r_i (t) \cdot ({\bf b}_i - {\bf b}_{i-1}) \right \rangle_{N_c}
\end{equation}
The function $C(t)$ is recast as:
\begin{equation}
\label{bonddisplcor2}
C(t) = \frac{1}{2} [ 1 - C_{bb}(t)] 
\end{equation}
To prove Eq.\ref{bonddisplcor2}, we notice that the difference of the displacements of the two adjacent monomers $i$ and $i+1$ is:
\begin{equation}
\Delta \mathbf r_i (t) - \Delta \mathbf r_{i+1} (t) = {\bf b}_i -  {\bf b}_i (t)
\label{tb_diffBond}
\end{equation}
Let us consider the sum in Eq.\ref{bonddisplcor}. Inserting Eq.\ref{tb_diffBond}  and setting ${\bf b}_M = {\bf b}_0 = 0$  yield:
\begin{eqnarray}
\label{tbPart0}\nonumber
& \displaystyle\sum_{i=1}^{M}  \Delta \mathbf r_i (t) \cdot \big({\bf b}_i - {\bf b}_{i-1}\big) = \\ [10pt]\nonumber
&\Delta \mathbf r_1 (t)\cdot {\bf b}_1 +  \Delta \mathbf r_2 (t) \cdot \big({\bf b}_2 - {\bf b}_1\big)+\ldots+\Delta 
\mathbf r_M (t)\cdot (-{\bf b}_{M-1}) = \\ [10pt]\nonumber
& \Big[ \Delta \mathbf r_1 (t) - \Delta \mathbf r_2 (t)\Big]\cdot {\bf b}_1 + \ldots + \Big[ \Delta \mathbf r_{M-1} (t) - 
\Delta \mathbf r_M (t)\Big]\cdot {\bf b}_{M-1} = \\ [10pt]\nonumber
& \Big[ {\bf b}_1 -  {\bf b}_1 (t) \Big]\cdot {\bf b}_1+\ldots+\Big[ {\bf b}_{M-1} -  {\bf b}_{M-1} (t) \Big]\cdot {\bf 
b}_{M-1} = \\ [10pt]\nonumber
& \displaystyle\sum_{i=1}^{M-1} {\bf b}_i^2 - \Big[{\bf b}_i (t) \cdot {\bf b}_i\Big] =\end{eqnarray}
\begin{equation}
\Big[M-1\Big]b^2 - b^2 \displaystyle\sum_{i=1}^{M-1} \frac{ {\bf b}_i (t) \cdot {\bf b}_i}{b^2}\label{tbPart3}
\end{equation}

Plugging Eq.\;\ref{tbPart3} into Eq.\;\ref{bonddisplcor} recovers Eq.\ref{bonddisplcor2} :
\begin{eqnarray}\nonumber
 C(t)&=& \left \langle  \frac{1}{2} - \frac{1}{2} \frac{1}{M-1} \sum_{i=1}^{M-1}  \frac{ {\bf b}_i (t) \cdot {\bf 
b}_i}{b^2}\right\rangle_{N_c}\\ [10pt]  \nonumber
&=& \frac{1}{2} \Big[ 1 - C_{bb}(t)\Big] 
\end{eqnarray}
By plugging Eq.\ref{bonddisplcor2} into Eq.\ref{tetaB_appnAPP}, one recovers Eq.\ref{tetaB_appn} .

\section{Short-chain limit of $\widetilde C(t)$}
\label{Appendix2}

We specialise $\widetilde C(t)$, Eq.\ref{tetaB_appn}, to short chains, i.e. we neglect the role of the chain entanglements. To this aim, one considers the Rouse gaussian theory of polymer dynamics which pictures the chains as 
"phantoms", i.e. perfectly crossable, and dissolved in a structureless environment \cite{DoiEdwards,Molin09}. The Rouse 
approximated expression of the { connectivity }  anisotropy function will be denoted as $\widetilde C^R(t)$ and defined by:
\begin{equation}
\label{CrHat}
\widetilde C^R(t) =  \kappa \, \frac{b}{2} \, \left [ \frac{3 \pi}{8 \, \langle \Delta r^2 (t)\rangle^R} \right ]^{1/2} \, \left [ 1 - C^R_{bb}(t) \right ]
\end{equation}
The above equation, with $\kappa=1$, is Eq. \ref{tetaB_appn} by taking into account that for gaussian displacements, one assumption of the Rouse theory \cite{DoiEdwards,Molin09},  $\langle \left|  \Delta \mathbf r_m (t)  \right|  \rangle = \sqrt{8/3 \pi} \, \langle \Delta r^2(t) \rangle^{1/2}$. {$\kappa $ is an adjustable parameter, independent of the chain length, to correct the approximations inherent in both the Rouse approach and the derivation of $\widetilde C(t)$. $C_{bb}^R(t)$ and $\langle \Delta r^2 (t)\rangle^R$ are the  expressions of $C_{bb}(t)$ and MSD in the framework of the Rouse theory, respectively \cite{DoiEdwards,Molin09}:
\begin{eqnarray}
C_{bb}^R(t) &=&    \frac{1}{(M-1)} \sum_{p=1}^{M-1} \phi^R_{p}(t) \label{Cr} \\
\langle \Delta r^2 (t) \rangle^R &=& \frac{b^2}{2 M} \left [  \frac{t}{\tau} +  \sum_{p=1}^{M-1} \frac{1-\phi^R_{p}(t)}{\sin^2(p\pi/2M)} \right ] \label{masdR}
\end{eqnarray}
where $\tau$ is a characteristic time, independent of the chain length, and $\phi^R_p(t)$ denotes the correlation function of the $p$-th Rouse mode \cite{DoiEdwards,Molin09}:
\begin{equation}
\phi^R_{p}(t)= \exp\left[-\sin^2(p\pi/2M) \,\frac{t}{ \tau}\right]
\label{eq:selfRouseprediction}
\end{equation}
Eq.\ref{Cr} and Eq.\ref{masdR} are derived by first expressing the monomer positions in terms of the Rouse orthogonal 
normal modes and then averaging the results over all the monomers  \cite{DoiEdwards,Molin09}.

\vspace{1cm}

\bibliography{biblio.bib}

\end{document}